\documentclass[12pt]{article}

\usepackage{epsfig}
\usepackage{amssymb}
\usepackage{a4}
\newcommand{\be}{\begin{equation}}
\newcommand{\ee}{\end{equation}}
\newcommand{\ba}{\begin{array}}
\newcommand{\ea}{\end{array}}

\newcommand{\bea}{\begin{eqnarray}}
\newcommand{\eea}{\end{eqnarray}}

\def\b#1{{\mathbb #1}}

\def\nn{\nonumber  \\}

\begin{document}

\title{Some explicit travelling-wave solutions  
of a perturbed sine-Gordon equation}

\author{ {\sc  Gaetano Fiore}
\\\\
Dip. di Matematica e Applicazioni, Universit\`a ``Federico II'',\\
 V. Claudio 21, 80125 Napoli\\
}

\date{}

\maketitle
\abstract{We present in closed form some special travelling-wave
solutions  (on the real line or on the circle) of a perturbed 
sine-Gordon equation. The perturbation of the equation consists
of a constant forcing term $\gamma$ and a linear
 dissipative term, and the equation  is used to describe the
Josephson   effect in the theory of superconductors and other
remarkable physical phenomena. We determine all travelling-wave solutions with
unit velocity (in dimensionless units). For $|\gamma|\!\le\! 1$ 
we find families of solutions that are all (except the obvious
constant one) manifestly unstable, whereas for $|\gamma|\!>\! 1$ 
we find families of stable solutions describing each an array of
evenly spaced kinks.}

\vfill
\noindent
Preprint 07-09 Dip. Matematica e Applicazioni, Universit\`a di Napoli.
To appear in the proceedings of the conference ``Mathematical Physics Models and Engineering Sciences'', Naples 22-23 June 2006, in honour 
of P. Renno's  for his 70$^{th}$ birthday.

\newpage

\section{Introduction and preliminaries}
The scope of this communication is the determination in closed
form of some special solutions of the class of partial differential equations 
\be                               \label{equation}
\varphi_{tt}-\varphi_{xx}
    + \sin\varphi+\alpha \varphi_t+\gamma =0
    \qquad x\in\b{R},
 \ee
parametrized by
constants $\alpha\!>\!0,\gamma\in \b{R}$, more precisely
the determination of the travelling-wave solutions 
$\varphi(x,t)=\tilde g(x\!-\!vt)$  with velocity $v=\pm 1$.

This equation (here written in dimensionless units) has been used to
describe with a good approximation a number of interesting physical phenomena,
notably Josephson effect in the theory of superconductors
\cite{Jos}, which is at the base \cite{BarPat82} of a large number of
advanced developments  both in fundamental research (e.g.
macroscopic effects of quantum physics, quantum computation) and in
applications to electronic devices (see e.g. Chapters 3-6 in
\cite{ChrScoSoe99}), or  more recently also the
 propagation of localized
 magnetohydrodynamic modes in plasma physics
\cite{Sco04}. The last two
 terms are respectively a dissipative and a forcing
one; the  sine-Gordon equation (sGe) is obtained by setting them
equal to zero.

\begin{figure}[ht]
\begin{center}
\epsfig{file=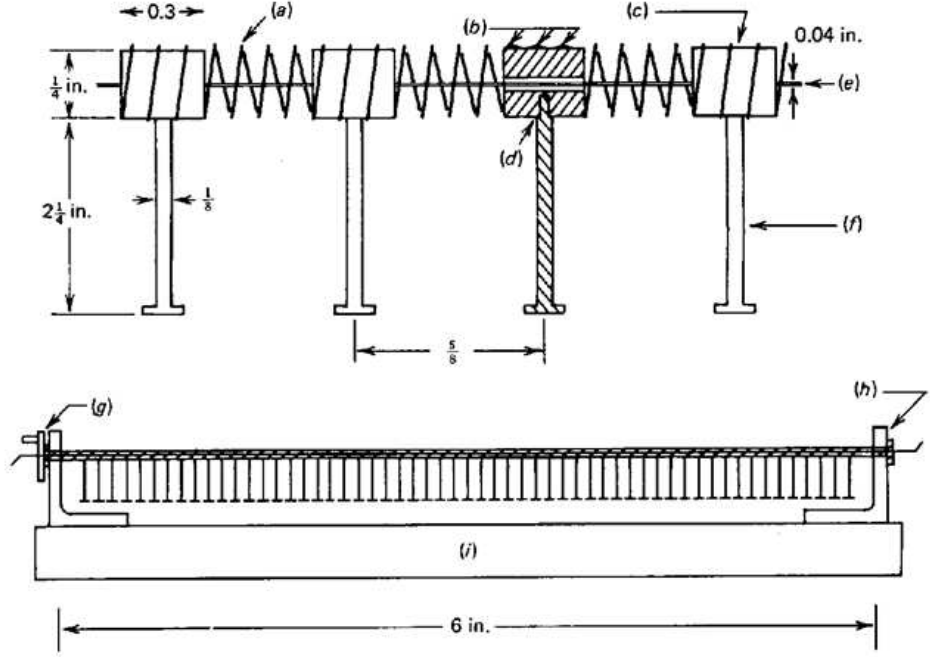 ,width=13truecm}
\caption{Mechanical model for the sine-Gordon equation. (a) Spring, (b)
solder, (c) brass, (d) tap and thread, (e) wire, (f) nail, (g) and (h) ball
bearings, (i) base (After A. C. Scott \cite{Sco70}, courtesy of A. Barone, see
\cite{BarPat82})}
\label{PendulaChain}
\end{center}
\end{figure}

The sGe describes also the dynamics of
the continuum limit of a sequence of neighbouring heavy pendula constrained
 to rotate around the same horizontal $x$-axis and coupled to each other
through a torque spring  \cite{Sco70}
(see fig. \ref{PendulaChain}); $\varphi(x,t)$
is the deviation angle from the lower vertical position  at time $t$ of the
pendulum having position $x$.
One can model also
the dissipative term $-\alpha \varphi_t$ of
(\ref{equation}) by immersing the pendula in a linearly viscous fluid, and the forcing term
$\gamma$ by assuming that a uniform, constant torque distribution is applied to
the pendula.
This mechanical analog allows a qualitative comprehension of the main
features of the solutions, e.g. of their instabilities.
The constant solutions of (\ref{equation}) exist only for $|\gamma|\le 1$ and
are, mod $2\pi$, 
\be
\varphi^s(x,t)\equiv - \sin^{-1}\gamma,
\qquad\qquad\varphi^u(x,t)\equiv  \sin^{-1}\gamma\!+\!\pi.
\ee 
If  $|\gamma|\!<\! 1$ the former is stable, the latter unstable, 
as they yield respectively local minima and maxima of the energy density
\be
h:=\frac {\varphi_t^2}2+
\frac {\varphi_x^2}2+\gamma\varphi-\cos\varphi.         \label{endensity}
\ee
In the mechanical analog they
respectively correspond to configurations with all pendula hanging down or
standing up. If $\gamma=\pm 1$ 
$\varphi^s=\varphi^u=\mp\pi/2\mbox{ (mod }2\pi)$, 
which is unstable because it is an inflection point for $h$.


\medskip In \cite{Fio05} we have performed a detailed analysis of 
travelling-wave solutions
of (\ref{equation}). We briefly recall the framework 
adopted there and some of the results.
Without loss of generality we can and shall assume $\gamma \ge 0$: if originally
$\gamma\!<\! 0$, we just need to replace $\varphi\to -\varphi$.
Moreover, space or time translations transform any solution into a
two-parameter family of solutions, so one can choose any of them as the
family representative element; for travelling-wave solutions
this reduces to translation of the only independent variable.
In agreement with the conventions adopted in \cite{Fio05},
we specify our travelling-wave Ansatz as follows:
\be
\xi:=\pm x- t,\qquad\quad
 \varphi(x,t)\!=\!g(\xi)\!-\!\pi.   \label{redef'}
\ee
Replacing the Ansatz in (\ref{equation}) one obtains the first order ordinary 
differential equation
\be
\alpha  g' = \gamma-\sin  g.  \label{equa}
\ee
We have already recalled the constant solutions. If $g'$ is not identically 
zero,  by integrating $d\xi=\alpha dg/(\gamma\!-\!\sin g)$ one finds
$$
\xi-\xi_0=\int\limits^{\xi}_{\xi_0}d\xi=
\alpha\int\limits^{
g}_{ g_0} \frac{ds}{\gamma-\sin s}
$$
separately in each interval in which $g'$ keeps its sign.
This allows to determine the solution implicitly, namely the inverse
$\xi(g)$. 

If $\gamma\!\le\! 1$, as $g$ approaches respectively $\sin^{-1}\!\gamma$ or
$\pi\!-\!\sin^{-1}\!\gamma$ (mod. $2\pi$) the denominator 
of the integrand goes to zero (linearly if $\gamma\!<\! 1$, quadratically if
$\gamma\!=\! 1$)
while keeping the same sign, and therefore the integral  diverges,
implying that the corresponding $\xi$ 
respectively goes either to $\pm \infty$, or to $\mp \infty$ \cite{Fio05}. In either case the range
of $\xi(g)$ is the whole $\b{R}$, implying that by taking the inverse
one obtains $g(\xi)$ already in all the domain.
If $\gamma\!>\! 1$ the denominator of the integrand is positive for all $s\in\b{R}$,
so that the solution $g$
is strictly monotonic and linear-periodic, i.e. the sum of a linear and a
periodic function, and
\be
g(\xi+\Xi)=  g( \xi)+2\pi,  \qquad\qquad\Xi:=
\alpha \int^{2\pi}_{0}\frac{ds}{\gamma-\sin s}.
\label{linper}
\ee
Denoting as  $\check\varphi^{\pm}$ the corresponding
solutions with $\xi\!:=\!\pm\! x\!-\! t$, by (\ref{redef'}) this implies
\be
\check\varphi^{\pm}(x\!+\!\Xi,t)=\check\varphi^{\pm}(x,t)\pm 2\pi.
\label{circle}  \ee
This behaviour is illustrated in fig. \ref{Photographs} by a picture
of the corresponding configuration for the mechanical model
of fig. \ref{PendulaChain}.

$\check\varphi^{\pm}$ can be interpreted 
also as solutions of (\ref{equation}) on
a circle of length $L=m \Xi$, for any $m\in\b{N}$. The integer $m$ parameterizes
different topological sectors: in the $m$-th the  pendula chain twists around
the circle $m$ times.

\section{Explicit travelling-wave solutions with unit velocity}

The purpose of this work is to determine in closed form
the travelling-wave solutions (\ref{redef'}) just described. 
We first transform eq. (\ref{equa}), with the help of the
identities (\ref{rationalize}), into
$$
4\alpha\frac{F'}{1\!+\!F^2}=\gamma-4\frac{F(1\!-\!F^2)}{(1\!+\!F^2)^2}
$$
by looking for $g$ in the form  $g=4 \tan^{-1}F$ and then into
\be
2\alpha y'=2y+\gamma(1\!+\!y^2)  \label{eqy}
\ee
by looking for $F$ in the form $F=y\!+\!\sqrt{1\!+\!y^2}$.
Note that diverging of $|y|$ at some point $\xi_0$
does not affect the continuity (and smoothness) of $g$ at $\xi_0$, 
even if the right limit is $\infty$
and the left one is $-\infty$, or viceversa:
$y\to\pm \infty$ respectively implies $F\to \infty, 0$
whence $g\to 0$ mod $2\pi$ in either case, which is compatible with a 
continuous $g$. 

Below we solve for $y(\xi)$ explicitly. Putting all redefinitions together,
we shall find solutions $\varphi$ through the formula
\be
 \varphi^{\pm}(x,t)\!=\!4 \tan^{-1}\left[y(\pm x\!-\! t)\!+\!
\sqrt{1\!+\!y^2(\pm x\!-\! t)}\right]
\!-\!\pi.   \label{redef"}
\ee

Only if $\gamma\le 1$ the solutions
$y_{\pm}=-\!\gamma^{-1}\pm \sqrt{\gamma^{-2}\!-\!1}$ of the second degree
equation $y^2\!+\!y2/\gamma\!+\!1=0$ are real and therefore give
(real) constant solutions $y(\xi)\equiv y_{\pm}$ of (\ref{eqy}), whence the
already mentioned constant solutions $\varphi^s,\varphi^u$
of (\ref{equation}). For nonconstant solutions (\ref{eqy}) is equivalent to
\be
d\xi= \frac{2\alpha}{\gamma} \frac{dy}{y^2+ \frac2{\gamma}y+1} \label{zuzu}
\ee
separately in each interval where $y'$ keeps its sign.
The discussion of  (\ref{zuzu}) depends now on the value of the
discriminant $\Delta=4/\gamma^2\!-\!4$ of the
equation $y^2\!+\!y2/\gamma\!+\!1=0$.

\medskip
If $\gamma\!<\!1$, then $\Delta>0$, $y_{\pm}$ are real and different 
and (\ref{zuzu}) can be written as
$$
d\xi= \frac{2\alpha}{\gamma} \frac{dy}{(y-y_+)(y-y_-)}
= \frac{\alpha} {\sqrt{1\!-\!\gamma^2}}
\left[\frac{dy}{y-y_+}-\frac{dy}{y-y_-}\right],
$$
which is integrated to give the two families of solutions
\be
y_1(\xi)=\frac{y_++y_- e^{A(\xi\!-\!\xi_0)} }{1+e^{A(\xi\!-\!\xi_0)} },
\qquad\qquad 
y_2(\xi)=\frac{y_+-y_- e^{A(\xi\!-\!\xi_0)} }{1-e^{A(\xi\!-\!\xi_0)} },
\ee
where $A:=(\sqrt{1\!-\!\gamma^2})\alpha^{-1}$
and $\xi_0$ is an integration constant. One easily checks that
$y_1',y_2'$ (and therefore also $g_1',g_2'$) 
are respectively negative-, positive-definite; and that
$\lim_{\xi\to \pm\infty}y_i(\xi)=
y_{\mp}$ for both $i=1,2$. Using formulae (\ref{F+}-\ref{F-}) 
shown in the Appendix
we thus find
$$
\lim_{\xi\to\infty}F\big[y_i(\xi)\big]=F(y_-)=\tan\theta, \qquad\qquad
\lim_{\xi\to-\infty}F\big[y_i(\xi)\big]=F(y_+)=\tan
\left(\frac{\pi}4\!-\! \theta\right).
$$
for both $i=1,2$, and mod $2\pi$ on one side a stricly
 decreasing $g_1(\xi)$ with
$$
\lim_{\xi\to -\infty} g_1=\pi- \sin^{-1}\gamma,\qquad\qquad
\lim_{\xi\to \infty} g_1= \sin^{-1}\gamma,
$$
and on the other a stricly increasing $g_2(\xi)$ with
$$
\lim_{\xi\to -\infty} g_2=\pi- \sin^{-1}\gamma,\qquad\qquad
\lim_{\xi\to \infty} g_2=2\pi+\sin^{-1}\gamma.
$$
As already noted, the singularity of $y_2$ at $\xi=\xi_0$ does not affect 
the continuity (and smoothness) of $g_2$.
Correspondingly, mod $2\pi$ 
\bea
&&\lim_{x\to \mp\infty} \varphi^{\pm}_1=- \sin^{-1}\gamma\equiv\varphi^s,\qquad\qquad
\lim_{x\to \pm\infty} \varphi_1^{\pm}= -\pi+\sin^{-1}\gamma\equiv\varphi^u,\\
&&\lim_{x\to \mp\infty} \varphi_2^{\pm}=- \sin^{-1}\gamma\equiv\varphi^s,\qquad\qquad
\lim_{x\to \pm\infty} \varphi_2^{\pm}=\pi+\sin^{-1}\gamma\equiv\varphi^u,
\eea
therefore $\varphi_1^{\pm},\varphi_2^{\pm}$ are unstable solutions, 
as noted in \cite{Fio05}.

\medskip
If $\gamma=1$, then $\Delta=0$, 
$y_{\pm}=-1$ and (\ref{zuzu})
can be written as
$$
d\xi= 2\alpha\:
\frac {dy}{(y+1)^2} = -2\alpha\:
d\!\left[\frac 1{y+1}\right],
$$
which is integrated to give
\be
y(\xi)=-\left[1+\frac{2\alpha} {\xi\!-\!\xi_0}\right]. 
\ee
This implies, with the help of (\ref{pi8}),
$$
\lim_{\xi\to \pm\infty} y(\xi)=-1, \qquad \Rightarrow \qquad
\lim_{\xi\to \pm\infty}F\big[y(\xi)\big]=\sqrt{2}\!-\!1=\tan \frac{\pi}8,
$$
whereas again
the singularity of $y$ at $\xi=\xi_0$ does not affect the continuity of $g$. 
As $y'$, and therefore also $F',g'$, are positive-definite, one finds mod $2\pi$
$$
\lim_{\xi\to -\infty} g=\frac{\pi}2,  \qquad\qquad
\lim_{\xi\to \infty} g=\frac{5\pi}2
$$
and, correspondingly, 
\be
\lim_{x\to \mp\infty} \varphi^{\pm}=-\frac{\pi}2, \qquad\qquad
\lim_{x\to \pm\infty} \varphi^{\pm}=\frac{3\pi}2;
\ee
also these $\varphi^{\pm}$ are  unstable, 
as noted in \cite{Fio05}.

\begin{figure}
\begin{center}
\epsfig{file=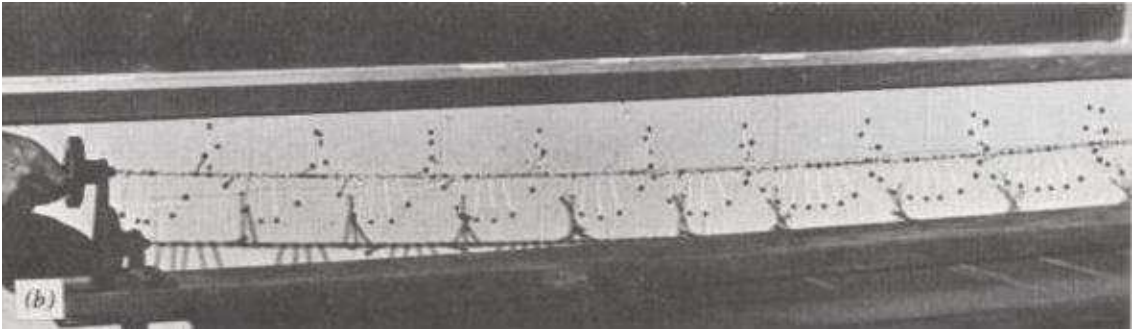,width=15truecm}
\caption{Photographs of the
mechanical model of fig. \ref{PendulaChain}: 
evenly spaced array of kinks (After A. C. Scott \cite{Sco70},
courtesy of A. Barone, see  \cite{BarPat82})} \label{Photographs}
\end{center}
\end{figure}

\medskip
Finally, if  $\gamma\!>\!1$, then $\Delta\!<\!0$,
$y_{\pm}$ are complex conjugate and
the denominator of (\ref{zuzu}) does not vanish for any value
of $y$. Setting $w:=(y\gamma\!+\!1)/\sqrt{\gamma^2\!-\!1}$
(\ref{zuzu}) can be written as
$$
d\xi= \frac{2\alpha} {\sqrt{\gamma^2\!-\!1}}\frac{dw}{1\!+\!w^2},
$$
which is integrated to give 
$\xi\!-\!\xi_0=2\alpha\tan^{-1}w/\sqrt{\gamma^2\!-\!1}$, whence
\be
y(\xi)=-\gamma^{-1}+ \frac 1{\sqrt{1\!-\!\gamma^{-2}}}
\tan\left[\frac {\sqrt{\gamma^2\!-\!1}}{2\alpha} (\xi-\xi_0)\right],
\ee
where $\xi_0$ is an integration constant. This is a
periodic function with period 
\be
\Xi:=\frac{2\pi\alpha}{\sqrt{\gamma^2\!-\!1}},
\ee 
and the latter
is also the period occurring in (\ref{linper}). In fact,
choosing $\xi_0=0$ for simplicity, we see that
as $\xi$ varies from $-\Xi/2$ to $\Xi/2$ $y(\xi)$ varies from $-\infty$ to $\infty$,
$F(\xi)$ varies  from $0$ to $\infty$, $g(\xi)$ varies  from $0$ to $2\pi$.
By continuity of $g$ (which again is not affected by
the singularity of $y$ at $\xi=\Xi(k\!+\!1/2)$ ($k\in\b{Z}$),
we thus find the behaviour (\ref{linper}).
The corresponding solutions $\check \varphi^{\pm}$ fulfill (\ref{circle}), 
describe  arrays
of evenly-spaced kinks (see fig. \ref{Photographs}) moving with velocity $\pm 1$
and are stable \cite{Fio05} (see also \cite{Sco69',ChuMcLSco73}).

\section*{Appendix}
We first recall the trigonometric identities
\be
\sin\! 2\alpha\!=\! \frac{2\tan\!\alpha}{1\!+\!\tan^2\!\alpha},
\quad\cos\! 2\alpha\!=\! \frac{1\!-\!\tan^2\!\alpha}{1\!+\!\tan^2\!\alpha}
\qquad\Rightarrow\qquad  
\sin\! 4\alpha\!=\! \frac{4\tan\!\alpha(1\!-\!\tan^2\!\alpha)}{(1\!+\!\tan^2\!\alpha)^2}. 
 \label{rationalize}
\ee

Given $\gamma\!\in\![0,1]$, let 
$\theta\!:=\!\frac 14 \sin^{-1}\!\gamma\in\![0,\frac{\pi}8]$. Then 
$\gamma=\sin 4\theta$, $\sqrt{1\!-\!\gamma^2}=\cos 4\theta$ and, using the
bisection formulae,
\bea
&&\sqrt{1\!+\!\sqrt{1\!-\!\gamma^2}}=\sqrt{1\!+\!\cos 4\theta}=
\sqrt{2}\cos 2\theta,\\
&&\sqrt{1\!-\!\sqrt{1\!-\!\gamma^2}}=\sqrt{1\!-\!\cos 4\theta}=
\sqrt{2}\sin 2\theta,
\eea
whence in turn
\bea
&&\sqrt{2}\!-\!\sqrt{1\!+\!\sqrt{1\!-\!\gamma^2}}=\sqrt{2}(1\!-\!
\cos 2\theta)=2\sqrt{2}\sin^2\theta,  \label{zaza}\\[8pt]
&&\sqrt{2}\!-\!\sqrt{1\!-\!\sqrt{1\!-\!\gamma^2}}=
\sqrt{2}(1\!-\!\sin\! 2\theta)=\sqrt{2}\left[1\!-\!\cos\!\left(\frac{\pi}2\!-\! 2\theta\right)\!\right]=2\sqrt{2}\sin^2\left(\frac{\pi}4\!-\! \theta\right).\qquad\quad
\eea
Hence, using also the sinus duplication formula, we end up with
\bea
F(y_+)&=&\frac{\sqrt{1\!-\!\sqrt{1\!-\!\gamma^2}}}{\gamma}
\left[\sqrt{2}\!-\!\sqrt{1\!-\!\sqrt{1\!-\!\gamma^2}}\right]=
\frac{4\sin 2\theta\sin^2\left(\frac{\pi}4\!-\! \theta\right)}
{\sin 4\theta} \nn[8pt]
&=&\frac{2\sin^2\left(\frac{\pi}4\!-\! \theta\right)}
{\cos 2\theta}=\frac{2\sin^2\left(\frac{\pi}4\!-\! \theta\right)}
{\sin\left(\frac{\pi}2\!-\! 2\theta\right)}=
\frac{\sin\left(\frac{\pi}4\!-\! \theta\right)}
{\cos\left(\frac{\pi}4\!-\! \theta\right)}=
\tan\left(\frac{\pi}4\!-\! \theta\right), \label{F+}\\[8pt]
F(y_-) &=&
\frac{\sqrt{1\!+\!\sqrt{1\!-\!\gamma^2}}}{\gamma}
\left[\!\sqrt{2}\!-\!\sqrt{1\!+\!\sqrt{1\!-\!\gamma^2}}\right]\!
=\frac{4\cos\! 2\theta\sin^2\!\theta}
{\sin\! 4\theta} 
=\frac{2\sin^2\!\theta}
{\sin\! 2\theta}=\tan\!\theta.\qquad\label{F-}
\eea

If we choose $\gamma\!=\!1$ in (\ref{zaza}) and  use the
sinus duplication formula we find as another consequence
\be
\sqrt{2}\!-\!1=2\sqrt{2}\sin^2\left(\frac{\pi}8\right)
=\frac{2\sin^2\left(\frac{\pi}8\right)}{\sin\left(\frac{\pi}4\right)}
=\frac{\sin\frac{\pi}8}{\cos\frac{\pi}8}=\tan\frac{\pi}8.\label{pi8}
\ee


\begin{thebibliography}{99}




\bibitem{BarPat82}
A. Barone, G. Patern\'o,
{\it Physics and Applications of the Josephson Effect},
 Wiley-Interscience, New-York, 1982; and references therein.

\bibitem{ChrScoSoe99}
P. l. Christiansen, A. C. Scott, M. P. Sorensen,
{\it Nonlinear Science at the Dawn of the 21st Century},
Lecture Notes in Physics 542, Springer, 2000.

\bibitem{ChuMcLSco73} See e.g.:  A. C. Scott, F. Y. F. Chu,
and  D. W. McLaughlin, {\it The soliton: a new concept
in applied science}, Proc. IEEE {\bf 61} (1973), 1443-1483.

\bibitem{DanDeaFio05}  A. D'Anna, M. De Angelis, G. Fiore, {\it Towards
soliton solutions of a perturbed sine-Gordon equation},
{\em Rend. Acc. Sc. Fis. Mat. Napoli}  {\bf LXXII} (2005), 95-110.
math-ph/0507005

\bibitem{Fio05} G. Fiore,
{\it Soliton and other travelling-wave solutions for a perturbed
sine-Gordon equation}, Preprint 05-49 Dip. Matematica e Applicazioni,
Univ.  ``Federico II'', e DSF/42-2005. math-ph/0512002.





\bibitem{Jos} Josephson B. D. {\it Possible new effects in superconductive
tunneling}, Phys. Lett. {\bf 1} (1962), 251-253; {\it The discovery of
tunneling supercurrents},  Rev. Mod. Phys. B {\bf 46} (1974), 251-254; and
references therein.













\bibitem{Sco69'}  A. C. Scott, {\it A nonlinear Klein-Gordon Equation}.
Am. J. Phys. {\bf 37} (1969), 52-61.

\bibitem{Sco70}  A. C. Scott, {\it Active and Nonlinear Wave Propagation
in Electronics}. Wiley-Interscience, New-York, 1970, Chapters 2,5.

\bibitem{Sco04}
J.L. Shohet, B.R. Barmish, H.K. Ebraheem, and A.C. Scott,
{\it The sine-Gordon equation in reversed-field pinch experiments},
Physics of Plasmas {\bf 11} (2004), 3877-3887.





\end{thebibliography}
\end{document}